\documentclass[usenatbib]{mn2e}
\input{psfig}
\newcommand{\galics}{{\sc galics}}
\newcommand{\momaf}{{\sc momaf}}
\newcommand{\ahop}{{\tt AdapatHOP}}

\newcommand{\blaizota}{{\sc momaf}}

\newcommand{\beq}{\begin{eqnarray}}
\newcommand{\eeq}{\end{eqnarray}}

\newcommand{\apj}{ApJ}

\newcommand{\apjl}{ApJL}
\newcommand{\aj}{AJ}
\newcommand{\mnras}{MNRAS}
\newcommand{\aap}{A\&A}

\newcommand{\avg}[1]{\langle{#1}\rangle}

\begin{document}
\title[{\sc galics} V -- Clustering in mock SDSS's] 
      {GalICS V : Low and high order clustering in mock SDSS's}
\author[Blaizot et al.]
       {J\'er\'emy Blaizot$^1$, 
	 Istv\'an Szapudi$^2$,
	 St\'ephane Colombi$^3$, 
	 Tam\'as Budav\`ari$^4$,\and
	 Fran\c{c}ois R. Bouchet$^3$, 
	 Julien E. G. Devriendt$^{5}$, 
	 Bruno Guiderdoni$^5$, \and
	 Jun Pan$^2$, Alex Szalay$^4$\\ 
	$^1$ Max-Planck-Institut f\"ur Astrophysik, Karl-Schwarzschild-Str. 1, 85741 Garching, Germany\\ 
	$^2$ Institute for Astronomy, University of Hawaii, 2680 Woodlawn Drive, Honolulu, HI 96822, USA\\
	$^3$ Institut d'Astrophysique de Paris, 98 bis boulevard Arago, 75014 Paris, France\\ 
	$^4$ Department of Physics and Astronomy, The Johns Hopkins University, Baltimore, MD 21218, USA\\
	$^5$ Centre de Recherche Astronomique de Lyon, 9 Avenue Charles Andr\'e, 69561 St-Genis-Laval Cedex, France}
\date{}
\maketitle
\label{firstpage}

\begin{abstract}
We use the GALICS hybrid model of galaxy formation
\citep{HattonEtal03} to explore the nature of galaxy clustering in the
local universe. We bring the theoretical predictions of our model into
the observational plane using the MOMAF software \citep{BlaizotEtal05}
to build mock catalogues which mimic SDSS observations. We measure low
and high order angular clustering statistic from these mock
catalogues, after selecting galaxies the same way as for observations,
and compare them directly to estimates from SDSS data. Note that we
also present the first measurements of high-order statistics on the
SDSS DR1. We find that our model is in general good agreement with
observations in the scale/luminosity range where we can trust the
predictions. This range is found to be limited (i) by the size of the
dark matter simulation used -- which introduces finite volume effects
at large scales -- and by the mass resolution of this simulation --
which introduces incompleteness at apparent magnitudes fainter than
$r\sim 20$.

We then focus on the small scale clustering properties of galaxies and
investigate the behaviour of three different prescriptions for
positioning galaxies within haloes of dark matter.  We show that
galaxies are poor tracers both of DM particles or DM sub-structures,
within groups and clusters. Instead, SDSS data tells us that the
distribution of galaxies lies somewhat in between these two
populations. This confirms the general theoretical expectation from
numerical simulations and semi-analytic modelling.
\end{abstract}

\section{Introduction}
Understanding galaxy biasing has become one of the most exciting
challenges of galaxy formation theories, especially due to the
overwhelming data sets that are being acquired at many wavelength,
e.g. with the Sloan Digital Sky Survey \citep[SDSS,][]{YorkEtal00},
and other large scale or deep surveys. Comprehension of galaxy biasing
can help us in using large scale structure (LSS) surveys to constrain
cosmological parameters. Or, the other way around, assuming that the
cosmology is known, galaxy clustering sets fundamental constraints on
models of galaxy formation. It is the second line that this paper
follows.

Two fundamentally different approaches are being used to investigate
galaxy clustering from a theoretical viewpoint. The first one consists
in running cosmological simulations that describe both the dark matter
and the baryonic components of the Universe
\citep[e.g.][]{PearceEtal99, CenOstriker00, PearceEtal01,
YoshikawaEtal01, WeinbergEtal04}. This method, although describing in
the most realistic manner the processes of galaxy formation in the
cosmological context, suffers from its computational expenses. As a
result, large scale clustering can only be explored at the price of
small scales, or, in other words, one has to chose between volume and
mass resolution. It however remains the only way to describe DM and
baryons in a fully consistent manner (up to the resolution limits).
The second approach gathers a large variety of implementations of the
so-called {\it halo model}. As is well illustrated by
\citet{PeacockSmith00}, the philosophy here is that galaxy
clustering stems from three ingredients only, namely, (i) halo
clustering properties, (ii) halo occupation distribution, and (iii)
spatial distribution of galaxies within haloes. While modelling the
spatial distribution of haloes has become routine with the increasing
number of $N$-body simulations, the way to populate these haloes with
galaxies is still a matter of debate. One can basically distinguish
two routes among the methods for populating DM haloes with
galaxies. The first one is based on biasing schemes : given the halo
mass, one uses phenomenological bias prescriptions to assign a number
of galaxies of given type and luminosity to a halo. Examples of this
so-called ``halo occupation distribution'' (HOD) method are
e.g. \citet{JingMoBoerner98}, \citet{PeacockSmith00},
\citet{SomervilleEtal01}, \citet{ScoccimarroEtal01},
\citet{ScoccimarroSheth02}, \citet{BerlindWeinberg02},
\citet{YangEtal04}. The second route uses semi-analytic models (SAMs)
of galaxy formation to produce a physically motivated distribution of
galaxies within haloes. The SAM can either be fed with semi-analytic
halo merger trees
\citep[e.g.][]{KauffmannNusserSteinmetz97,BensonEtal00,BensonEtal01}
or merger trees directly extracted from cosmological DM simulations
\citep[e.g.][]{KauffmannEtal99a,HellyEtal03a,HattonEtal03}. In all
cases, the spatial distribution of haloes is taken from N-body
simulations. 

The last unknown in the framework of the halo-model is then the
spatial distribution of galaxies within the haloes they populate.
Much work has recently been done to understand the nature of this
distribution relative to the distribution of DM and it now seems clear
that galaxies sample sub-haloes within clusters in a non-trivially
biased manner \citep[e.g.][]{SpringelEtal01, GaoEtal04,
NagaiKravtsov05}. The bias arises because sub-haloes are stripped much
more efficiently than the galaxies they harbour as they orbit within the main halo's
potential well, which gives rise to a steeply decreasing mass-to-light
ratio inwards the halo. The HOD and SAM routes then again differ. On
the one hand, the HOD formalism distributes galaxies as a function of
an instantaneous view of the DM distribution. It is thus not suited to
describe the above evolutionary process and HOD implementations
usually assume that galaxies are distributed as the DM particles
within each halo, with the exception of the most massive galaxy which
is forced to lying at the centre of its host halo. Note that
\citet{GaoEtal04} suggest that this is a very good approximation,
although \citet{NagaiKravtsov05} find somewhat different results. On
the other hand, SAMs generally attempt to predict the galaxy spatial
distribution with a more or less detailed modelling of the dynamical
processes that shape it. This either involves semi-analytic
prescriptions that describe dynamical friction and how halo mergers
affect galaxy orbits \citep[e.g.][]{HattonEtal03}, or DM-based
treatments in which galaxies typically follow the most bound particle
of the halo in which they were formed \citep[e.g.][]{KauffmannEtal99a,
KauffmannEtal99b,DiaferioEtal99,DiaferioEtal01,MathisEtal02,
MathisWhite02,HellyEtal03a} or even DM sub-haloes\footnote{Because of
finite mass resolution and efficient tidal stripping of
sub-structures, the sub-haloes harbouring galaxies cannot all be
followed in practice. A proxy is then necessary : when a sub-halo
disappears, the galaxy it contained follows the most bound particle of
this sub-halo, identified before the halo vanishes.}
\citep{SpringelEtal01,DeLuciaKauffmannWhite04,GaoEtal04}.

The objective of this paper is to improve on previous theoretical
studies of galaxy clustering in the following directions. {\it First},
we use the state-of-the-art \galics{} model of galaxy formation
\citep{HattonEtal03} to populate DM haloes from a cosmological
$N$-body simulation. This model describes galaxy formation with
semi-analytic prescriptions applied within halo merger trees extracted
from that DM $N$-body simulation, and thus provides us with a
physically motivated halo occupation distribution (HOD). {\it Second},
we use the \momaf{} software \citep[][ hereafter
\momaf{}]{BlaizotEtal05} to construct mock catalogues that mimic the
SDSS early data release and DR1, both in geometry and photometric
selection. These mock catalogues enable us to carry out a direct
comparison of {\it angular} galaxy clustering statistics with those
observed in the SDSS. Note that comparisons of hybrid models and
observations have already been performed in the ``observational
plane'' in the past \citep{DiaferioEtal99,MathisEtal02}. {\it Third},
we extend the comparison to high order statistics such as the 3- and
4-point angular correlation functions.  {\it Fourth}, we investigate
how clustering statistics are affected by the spatial distribution of
galaxies within haloes, that is, what does the SDSS data tell us about
the small scale distribution of galaxies. To this end, we compare the
results obtained with three different galaxy distributions~: (i) the
one predicted by the ``progenitor position interpolation'' scheme
implemented in the standard version of \galics{}, (ii) one where
galaxies follow dark matter density within haloes, and (iii) one where
galaxies are distributed as DM substructures.  {\it Finally}, as part
of the \galics{} series, a side-output of this paper is the validation
of the combined tools \galics{} and \momaf{} concerning their ability
to predict spatial and angular clustering of galaxies in a more
general context, i.e. in the framework of forthcoming extra-galactic
surveys. This is particularly meaningful since all the data used in
this paper are available from the \galics{} web-page\footnote{{\tt
http://galics.cosmologies.fr/}}, in the form of a relational database
\citep[see][]{BlaizotEtal05}.

The paper is organised as follows. In Sec. \ref{sec:model} we review
the characteristics of \galics{} and \momaf{} which are relevant to the
present study. In Sec. \ref{sec:w} we discuss the two-point angular
correlation. In Sec. \ref{sec:Sn} we discuss higher order clustering
statistics. We discuss our results and conclude in Sec. \ref{sec:conclu}.

\section{Simulation and mock catalogues} \label{sec:model}
\galics{} is a hybrid model of galaxy formation which combines
cosmological DM simulations with a semi-analytic description of
baryonic processes. The model is fully described in
\citet{HattonEtal03}, and the version we use here is the same as that
used in the previous papers of the \galics{} series
\citep{HattonEtal03, DevriendtEtal05, BlaizotEtal04}. We briefly
remind the main ingredients in Secs. \ref{sec:dm} and
\ref{sec:baryons}. We have been lead to change our prescription for
positioning galaxies within haloes, since \citet{HattonEtal03}. We
explain our new prescription in Sec. \ref{sec:pos} \citep[also
read][]{LanzoniEtal05}. In this latter section, we present alternative
positioning schemes that we will explore in the following sections.

Eventually, \galics{}' outputs are turned into mock catalogues using
\momaf{}, as explained in Sec. \ref{sec:mocks}. We check in this
latter section that the basic properties of these mock catalogues,
i.e. number counts and red-shift distributions, are in agreement with
SDSS data.

\subsection{Dark matter} \label{sec:dm}
The cosmological N-body simulation \citep{Ninin99} we use throughout
this paper assumes a flat Cold Dark Matter cosmology with a
cosmological constant ($\Omega_m=1/3$, $\Omega_\Lambda = 2/3$), and a
Hubble parameter $h=H_0/[100$ km s$^{-1}$ Mpc$^{-1}] =0.667$. The
initial power spectrum was taken to be a scale-free ($n_s = 1$) power
spectrum evolved as predicted by \citet{BardeenEtal86} and normalised
to the present-day abundance of rich clusters with $\sigma_8 = 0.88$
\citep{EkeColeFrenk96}. The simulated volume is a cubic box of side
$L_b = 100 h^{-1}$Mpc, which contains $256^3$ particles, resulting in
a particle mass $m_p = 8.272\times 10^9 M_\odot$ and a smoothing
length of 29.29 kpc. The density field was evolved from $z=35.59$ to
present day, and we out-putted about 100 snapshots spaced
logarithmically with the expansion factor.

In each snapshot, we identify halos using a friend-of-friend (FOF)
algorithm \citep{DavisEtal85} with a linking length parameter $b=0.2$,
only keeping groups with more than 20 particles. At this point, we
define the mass $M_{FOF}$ of the group as the sum of the masses of the
linked particles, and the radius $R_{FOF}$ as the maximum distance of
a constituent particle to the centre of mass of the group. We then fit
a tri-axial ellipsoid to each halo, and check that the virial theorem
is satisfied within this ellipsoid. If not, we decrement its volume
until we reach an inner virialised region. From the volume of this
largest ellipsoidal virialised region, we define the virial radius
$R_{vir}$ and mass $M_{vir}$ . These virial quantities are the ones we
will use later to compute the cooling of the hot baryonic component.
Once all the halos are identified and characterised, we build their
merger history trees following all the constituent particles from
snapshot to snapshot.

\subsection{Lighting up haloes} \label{sec:baryons}
The fate of baryons within the halo merger trees found above is
decided according to a series of prescriptions which are either
theoretically or phenomenologically motivated. The guideline -- which
is similar to other SAMs -- is the following. Gas is shock-heated to
the virial temperature when captured in a halo's potential well. It
can then radiatively cool onto a rotationally supported disc, at the
centre of the halo. Cold gas is turned into stars at a rate which
depends on the dynamical properties of the disc. Stars then evolve,
releasing both metals and energy into the interstellar medium (ISM),
and in some cases blowing part of the ISM away back into the halo's
hot phase. When haloes merge, the galaxies they harbour are gathered
into the same potential well, and they may in turn merge together,
either due to fortuitous collisions or to dynamical friction. When two
galaxies merge, a ``new'' galaxy is formed, the morphological and
dynamical properties of which depend on those of its
progenitors. Typically, a merger between equal mass galaxies will give
birth to an ellipsoidal galaxy, whereas a merger of a massive galaxy
with a small galaxy will mainly contribute to developing the massive
galaxy's bulge component. The Hubble sequence then naturally appears
as the result of the interplay between cooling -- which develops discs
-- and merging and disc gravitational instabilities -- which develop
bulges.

Keeping track of the stellar content of each galaxy, as a function of
age and metalicity, and knowing the galaxy's gas content and chemical
composition, one can compute the (possibly extincted) spectral energy
distribution (SED) of each galaxy. To this end, we use the {\sc
stardust} model \citep{DevriendtGuiderdoniSadat99} which predicts the
SED of an obscured stellar population from the UV to the sub-mm.

The above modelling of galaxy formation provides us with a physically
motivated HOD : it tells us how many galaxies one expects in each halo
along with the properties of these galaxies. It also predicts the
dispersion (and higher orders) of the HOD, as a result of each halo's
individual formation history. In the \galics{} model, the number of
galaxies that populate a halo results from basically three
ingredients. {\it First}, gas cools in haloes massive enough compared
the the IGM temperature. This is the source term and produces one
(central) galaxy per (massive) halo. {\it Second}, galaxies gather in
the same structures when haloes merge. This is the only way to get
more than one galaxy per halo, and tends to yield a number of
satellite galaxies proportional to halo mass at high masses. {\it
Third}, galaxy-galaxy mergers are the only sink term (regardless
selection effects). In a paper in preparation, we show that the HOD
predicted by \galics{} is in good agreement with results from a
smoothed-particle hydrodynamics cosmological simulation, suggesting
that the three above ingredients and their implementation properly
capture the physics that shape the HOD.

\subsection{Galaxy positions} \label{sec:pos}
The position $\vec{p}_g$ of a galaxy in the simulation volume can be
written as $\vec{p}_g = \vec{p}_h + \delta\vec{p}$, where $\vec{p}_h$
is the position of the centre of mass of the host halo, and
$\delta\vec{p}$ the position of the galaxy within this halo. While the
positions of haloes are well known from the DM simulation, the spatial
distribution of galaxies inside their host haloes is not described by
DM-only simulations. One thus needs a model to predict each galaxy's
$\delta\vec{p}$. In this paper, we investigate the effect of three
such models on the clustering properties of galaxies~: (i) the
``progenitor position interpolation'' (PPI) implemented in the
standard version of \galics{}, (ii) a scheme in which galaxies follow DM
within haloes (FOF), and (iii) a model in which galaxies are positioned on
DM substructures (SUB). Seeing which scheme the SDSS data prefer will
hopefully help us understand how galaxies are distributed within DM
haloes.

\begin{figure}
\begin{center}
\psfig{file=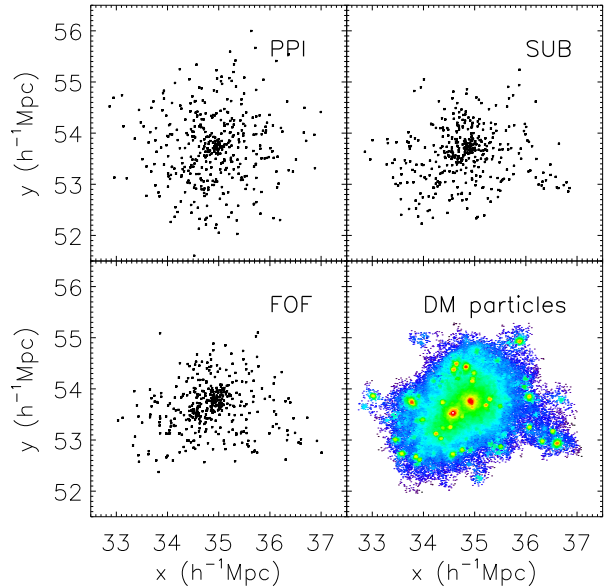,width=0.48 \textwidth}
\end{center}
\caption{Projected positions of galaxies (dots) in a $8.3\times
10^{14}$ M$_{\odot}$-halo containing 397 galaxies. The upper-left
(resp. upper-right, lower-left) panel shows the PPI (resp. SUB, FOF)
galaxy distribution. The lower-right panel shows the distribution of dark-matter particles, for comparison. The virial radius of this halo is $1.6h^{-1}$Mpc.}
\label{fig:positions}
\end{figure}

\vskip 0.2cm
\noindent {\bf PPI} -- Because of the spherical symmetry assumption
made in \galics{}, a galaxy's position in our model is described only
by its orbital radius.  We model two processes that can affect a
galaxy's distance to its halo's centre~: (i) dynamical friction brings
galaxies to the centre, and (ii) halo mergers heavily perturbate
galaxies' orbits. Because of the frequent mergers, it is the latter
process that mostly determines the galaxy distribution.  When two
haloes merge, the positions of the galaxies within the descendent halo
are obtained with an interpolation of the progenitor's positions using
their velocities.  In this paper, we use a new prescription to
reposition galaxies after halo mergers, which is a modified version of
that described in \citet{HattonEtal03}, designed to better take into
account the difference in masses of the merging haloes
\citep[see][]{LanzoniEtal05}. In practice, the displacement distance
$R_j$ of \citet[][eq. 5.1]{HattonEtal03} is now multiplied by a factor
$(1-M_{\tt prog}/M_{\tt son})$, where $M_{\tt prog}$ is the mass of
either progenitor, and $M_{\tt son}$ the mass of the descendent. In
this way, when a small halo merges with a much more massive one,
galaxies' orbits in the massive halo will change very little, whereas
galaxies' orbits in the small halo will change as before. The effect
of this prescription is to yield more concentrated galaxy
distributions than the original prescription did. Note however that
the overall properties of galaxies are almost identical to those
presented in \citet{HattonEtal03} and \citet{BlaizotEtal04}.  An
example of the galaxy distribution predicted by the ``PPI'' model is
shown in the upper-left panel of Fig. \ref{fig:positions}, for a halo
of mass $8.3\times 10^{14}$ M$_{\odot}$, containing 397 galaxies.

\vskip 0.2cm
\noindent {\bf FOF} -- The second positioning scheme explored in this
paper, and hereafter called ``FOF'', consists in placing galaxies on
random particles of the halo they belong to, with the exception of the
most massive galaxy which is forced to lie at the centre of mass of
its halo. This is done as a post-treatment of the \galics{} outputs
and has thus no effect on the physical properties of modelled
galaxies. For the same reason, though, the positions of {\it satellite}
galaxies within haloes are not related to their physical properties.
The resulting distribution is illustrated in the lower-left panel of
Fig. \ref{fig:positions}.  Several qualitative differences can be
noticed with the PPI distribution : (i) the shape of the distribution
is more complex (two cores, etc.), (ii) it is much more concentrated
near the core(s). Note that this FOF distribution is the one most
commonly used in HOD implementations.

\vskip 0.2cm
\noindent {\bf SUB} -- The third positioning prescription we explore
consists in placing galaxies on top of DM sub-structures (this will
hereafter be referred to as ``SUB''). In this case, we assign galaxies
to sub-structures as a function of their masses~: more massive
galaxies go to more massive sub-structures.  As a result, the most
massive galaxy of a halo naturally ends up at the centre of mass of
this halo.
This procedure relies on the assumption that the mass of a
sub-structure roughly scales with that of the galaxy it contains. This
assumption is definitely questionable, since sub-structures are much
more efficiently tidally stripped -- while orbiting within the main
halo -- than the galaxies they harbour
\citep[e.g.][]{SpringelEtal01,DiemandMooreStadel04,NagaiKravtsov05}. We
thus expect our procedure to induce a significant depletion of
identified galaxies in the cores of massive haloes. This should
definitely have some impact on the measurement of the two-point
correlation function and higher order statistics at small scales. In
particular, we expect that the SUB scheme will lead to an
under-estimate of the small-scale clustering signal, with respect to
what would be found with a full hydro-dynamical treatment as in
\citet{NagaiKravtsov05}. Still, the exercise is interesting because
our SUB and FOF schemes are expected to closely bracket the ``true''
distribution of galaxies.

Also note that in practice, the number of substructures within a halo
may differ from the number of SAM galaxies it contains. This is an
issue when there are less sub-structures than galaxies. In this case,
the extra (low-mass) galaxies are given the positions of random dark
matter particles as in the FOF scheme. Fortunately, for our SDSS mock
catalogues, such miss-identifications are rare enough, as shown in
Appendix \ref{sec:HOP}. The upper-right panel of
Fig. \ref{fig:positions} shows the ``SUB'' distribution of galaxies in
the same halo as before. This distribution lies somewhat in between
the FOF and PPI pictures. The identification of substructures is done
using the \ahop{} code \cite{AubertPichonColombi04}, as described in
Appendix \ref{sec:HOP}.

\vskip 0.2cm

Finally, note that in the 3 different positioning schemes, we do not allow for
galaxies to overlap, i.e. we impose a minimum distance between
galaxies of twice their sizes.

\subsection{Mock catalogues} \label{sec:mocks}
We use the {\it random tiling} technique described in \blaizota{} to
build mock observations from the redshift outputs of \galics{}. We mimic the
SDSS early data release by constructing catalogues of $2.5\times 90$
square degrees, limited in apparent magnitude at $r = 22$. As
explained in \blaizota{}, several different observing cones can be
generated from the same set of outputs of \galics{}, by changing
either the line of sight or the seed for the random tiling. We build
20 cones with seeds and lines-of-sight chosen randomly for each
positioning scheme. 
These 20 cones allow us to infer some estimate of the dispersion in
clustering measurements, that is, the cosmic variance associated to
our mock catalogues.  However, given the rather small size of the
simulation box, $100 h^{-1}$ Mpc on a side, this estimate is likely to
be biased and has to be taken as a lower boundary on the cosmic errors.

In Table \ref{tab:mocks}, we give some geometrical characteristics of
our mock catalogues. The first line gives the median red-shift ($z_{\tt
med}$) of each apparent-magnitude selection. The second and third
lines give the angular size ($\theta_b$) of our simulated volume at
$z_{\tt med}$ and the corresponding number of boxes required to fill
the observing cone in its largest dimension ($N_t \times \theta_b \sim
90 \deg$). 
Both these quantities give an idea of the importance of finite volume
and replication effects, which tend to reduce the amplitude of the
$N$-point correlations functions as well as that of the measured
cosmic variance on their estimates (see \citet{BlaizotEtal05} for a
thorough discussion of these effects).
The fourth and fifth lines give similar quantities, this time along
the line-of-sight. The sixth and seventh lines give the completeness
limits at $z_{\tt med}$ in terms of absolute rest-frame magnitudes in
the $r$-band ($r_{\tt res}$) and in the $I$-band ($I_{\tt
res}$). Fainter than these limits, our sample of galaxies is
incomplete due to resolution effects : we miss part of the galaxies
because they would lie in unresolved DM haloes. The last line of Table
\ref{tab:mocks} gives the observer-frame absolute magnitude
corresponding to the faint boundary of the selection at $z_{\tt med}$
in each magnitude bin. This magnitude should be compared to $r_{\tt
res}$ at low red-shifts and to $I_{\tt res}$ at higher
red-shifts. Comparison then tells us whether the sample of galaxies we
select with the apparent-magnitude cut is complete. As can be seen
from the two last columns of Table \ref{tab:mocks}, our samples of
galaxies become incomplete faint-wards $r \sim 20$.

\begin{table*}
\begin{center}
\begin{minipage}{15.5cm} \def\footnoterule{}
\begin{center}
\begin{tabular}{lcccc} 
\hline\hline 
                              & $18<r<19$       & $19<r<20$     & $20<r<21$      & $21<r<22$ \\
\hline
$z_{\tt med}$\footnote{Median red-shift of galaxies in each magnitude bin, as shown by the vertical lines in Fig. \ref{fig:zdist}.}
                              & 0.22            & 0.31          & 0.41           & 0.55      \\
$\theta_b$\footnote{Angular size of the simulated volume at $z_{\tt med}$, in degrees.}
                              & 11.2            & 8.77          & 7.34           & 6.26      \\
$N_t$\footnote{Number of boxes tiled across the line of sight, in the direction where the observing cone is 90 deg. wide.}
                              & 8               & 10            & 12             & 14        \\
$d_c(2z_{\tt med})$\footnote{Co-moving distance from the observer to
                              $2\times z_{\tt med}$, in $h^{-1}$Mpc.}
                              & 1170            & 1567          & 1957           & 2429      \\
$N_r$\footnote{Number of boxes tiled along the line of sight.}
                              & 12              & 16            & 20             & 25        \\
\hline
$r_{\tt res}$\footnote{Completeness magnitude limit in the SDSS $r$ band (rest-frame).}
                              & -20.0           & -20.5         & -20.6          & -20.3     \\
$I_{\tt res}$\footnote{Completeness magnitude limit in the $F814W$ band from HST (rest-frame).}
                              & -20.5           & -20.8         & -20.9          & -20.6     \\
$M_r$\footnote{Absolute (observer-frame) magnitudes corresponding to the fainter boundary of each apparent magnitude bin, at the corresponding median red-shift.}
                              & -21.3           & -21.1         & -20.8          & -20.6     \\
\hline
\end{tabular}
\end{center}
\end{minipage}
\caption{Summary of the limitations of our simulation, in terms of
volume (first five rows), and mass resolution (three last rows). The
angular size of our simulation allows us to probe clustering up to
scales ranging from $\sim 1$ to $\sim 0.6$ degrees from the brightest
to the faintest apparent magnitude bins. Moreover, the mass resolution
guarantees that our samples of galaxies are complete in the two
brightest magnitude bins, while we certainly miss part of the galaxies
at fainter fluxes.\label{tab:mocks}}
\end{center}
\end{table*}

\begin{figure}
\begin{center}
\psfig{file=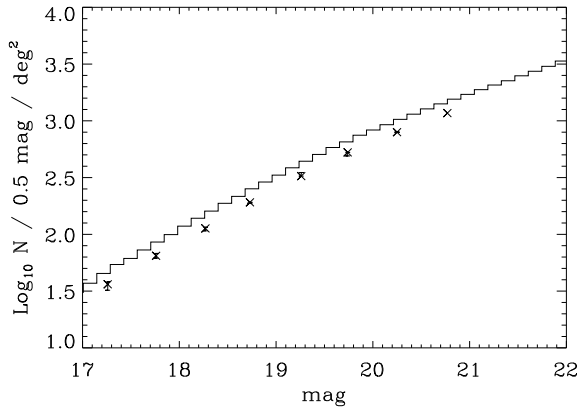,width=0.48 \textwidth}
\end{center}
\caption{Comparison of number counts from \galics{} (solid histogram)
with those from the SDSS early data release (crosses with error bars,
taken from \citet{YasudaEtal01}), in the $r$ filter.}
\label{fig:counts}
\end{figure}

Before approaching clustering statistics, one should first check that
one-point statistics (number counts and red-shift distributions) are
in agreement with the data. In Fig. \ref{fig:counts}, we show the
comparison of \galics{} counts in the $r$ band (solid line) with the
SDSS observations (crosses with error bars) taken from
\citet{YasudaEtal01}. These counts were measured on one mock SDSS
stripe. Our model slightly over-estimates the observed counts at all
magnitudes, of $\sim$0.1 dex in number or $\sim 0.2$ mag in
magnitudes.  The reasons for this over-estimate are not obvious. They
are partly due to an over-estimate of the present-day luminosity
function, and possibly to a slightly wrong redshift evolution
(although see discussion of 2dF $N(z)$ in \momaf{}). The important
point here is that the counts match observations well enough for our
purposes. We indeed show in Sec. \ref{sec:w_dmag} that such a small
error in the number counts does not affect our conclusions concerning
clustering.

\begin{figure}
\begin{center}
\psfig{file=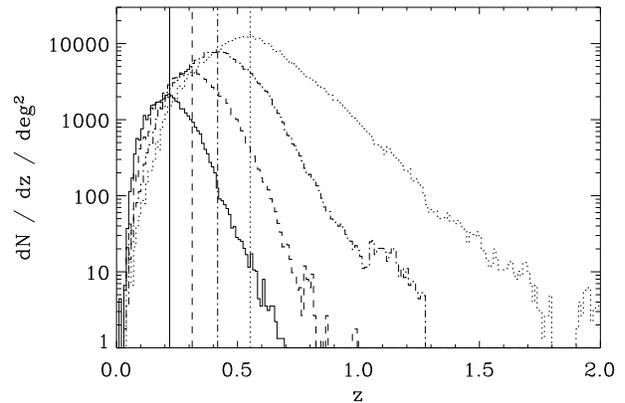,width=0.48 \textwidth}
\end{center}
\caption{Red-shift distributions of modelled galaxies selected in four
apparent-magnitude bins. The solid line (resp. dashed, dot-dashed,
dotted) corresponds to $18<r<19$ (resp. $19<r<20$, $20<r<21$,
$21<r<22$).}
\label{fig:zdist}
\end{figure}

In Fig. \ref{fig:zdist} we show the red-shift distributions of
modelled galaxies selected in four apparent-magnitude bins (hereafter
``standard'' magnitude bins). The solid line shows the red-shift
distribution of galaxies with apparent $r$ magnitude between 18 and
19, the dashed line is for $19<r<20$, the dot-dashed line for
$20<r<21$, and the dotted line for $21<r<22$. The median red-shifts of
each sample are respectively $z_{\tt med} = 0.22$, $0.31$, $0.41$,
$0.55$, as indicated with the vertical lines in
Fig. \ref{fig:zdist}. The median red-shift of the brightest bin is
consistent with $z_{\tt med} = 0.18$ given by \citet{ConnollyEtal02}.
Again, these redshift distributions were obtained from a single mock
catalogue. The (small) high-$z$ bump seems to be a general trend of
our model (it appears in most of our 20 mocks), and not due to a
particular super-structure. We have checked that this anomaly has no
effect on our clustering estimates by computing $w(\theta)$ with and
without galaxies in the high-$z$ tails : results are
undistinguishable.

\section{Two-point correlation function} \label{sec:w}
In this section, we first show that our clustering results do not
depend much on the uncertainty in the counts. Then, we present the
angular correlation function (ACF) we obtain with the PPI scheme for
positioning galaxies, and discuss it's agreement with SDSS
data. Finally, we explore how the three positioning schemes affect the
ACF at small scales.

\subsection{ACF estimate and robustness} \label{sec:w_dmag}
We compute the ACF $w(\theta)$ using the estimator proposed by
\citet{LandySzalay93}. However, instead of counting pairs, we use a
fast Fourier transform (FFT) scheme which is much faster when the
number of galaxies is large \citep{SzapudiPrunetColombi01}. This
method requires one to project the apparent galaxy density onto a
grid, the cell-size of which sets a lower limit to the scales one can
probe. We therefore project each mock SDSS strip on rectangular grids
of $16384 \times 455$ cells, which correspond to cells of size $\sim
20$ arc-seconds.

We do not attempt to correct for the integral constraint, as
\citet{ScrantonEtal02} showed that it is negligible. Moreover, because
we mimic the geometry of the SDSS, we are affected by the same
integral constraint as observations. A direct comparison of both raw
estimates then makes more sense. 
Also, we do not estimate errors analytically, as they are in principle
fully contained in the dispersion of our measurements among the 20
mock catalogues.  Remember however that we expect this dispersion to
give a lower bound on the errors rather than a true estimate, due to
the finite size of the simulation box.

\begin{figure}
\begin{center}
\psfig{file=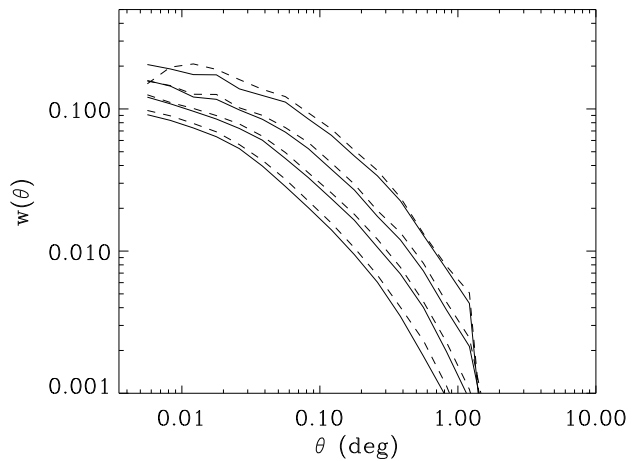,width=0.48 \textwidth}
\end{center}
\caption{Angular correlation functions of galaxies in different
apparent-magnitude bins 
using the PPI scheme to locate galaxies. 
The solid lines show the ACF of galaxies with
$18<r<19$, $19<r<20$, $20<r<21$ and $21<r<22$, from top to bottom. The
dashed lines show the ACFs of galaxies with $17.8<r<18.8$,
$18.8<r<19.8$, $19.8<r<20.8$ and $20.8<r<21.8$, from top to bottom. An
uncertainty of 0.2 mag translates in very little changes of
$w(\theta)$.}
\label{fig:w_dmag}
\end{figure}

As mentioned earlier, the counts from \galics{} slightly differ from
those given by \citet{YasudaEtal01}. The significance of this
discrepancy is not very clear and could possibly be due to
e.g. definition of magnitudes, or pollution by stars
\citep{YasudaEtal01,ScrantonEtal02}. A full treatment of photometric
errors is however beyond the scope of this paper. Instead, we show
that our results are not very sensitive to apparent magnitude
uncertainties. To do this, we use a single mock catalogue to compute
the angular correlation function for galaxies in the four standard
magnitude bins ($18<r<19$, $19<r<20$, $20<r<21$, $21<r<22$) and in
magnitude bins shifted by $-0.2$ mag ($17.8<r<18.8$, $18.8<r<19.8$,
$19.8<r<20.8$, $20.8<r<21.8$). This shift is about what is required
for our model counts to fit the SDSS counts. We show the results in
Fig. \ref{fig:w_dmag}
for one of the PPI catalogues. 
The solid lines correspond to the standard magnitude bins, and the
dashed ones to the shifted magnitudes. Naturally, we find that
brighter galaxies (the shifted bins) are more clustered than fainter
ones.
However, the shape of the correlation function is not affected, and
the difference in amplitude is very small. Similar results would be
obtained with the FOF and SUB schemes to position galaxies. Our
conclusions are thus robust in this prospect.

\subsection{Results for the PPI scheme}
\begin{figure*}
\begin{center}
\psfig{file=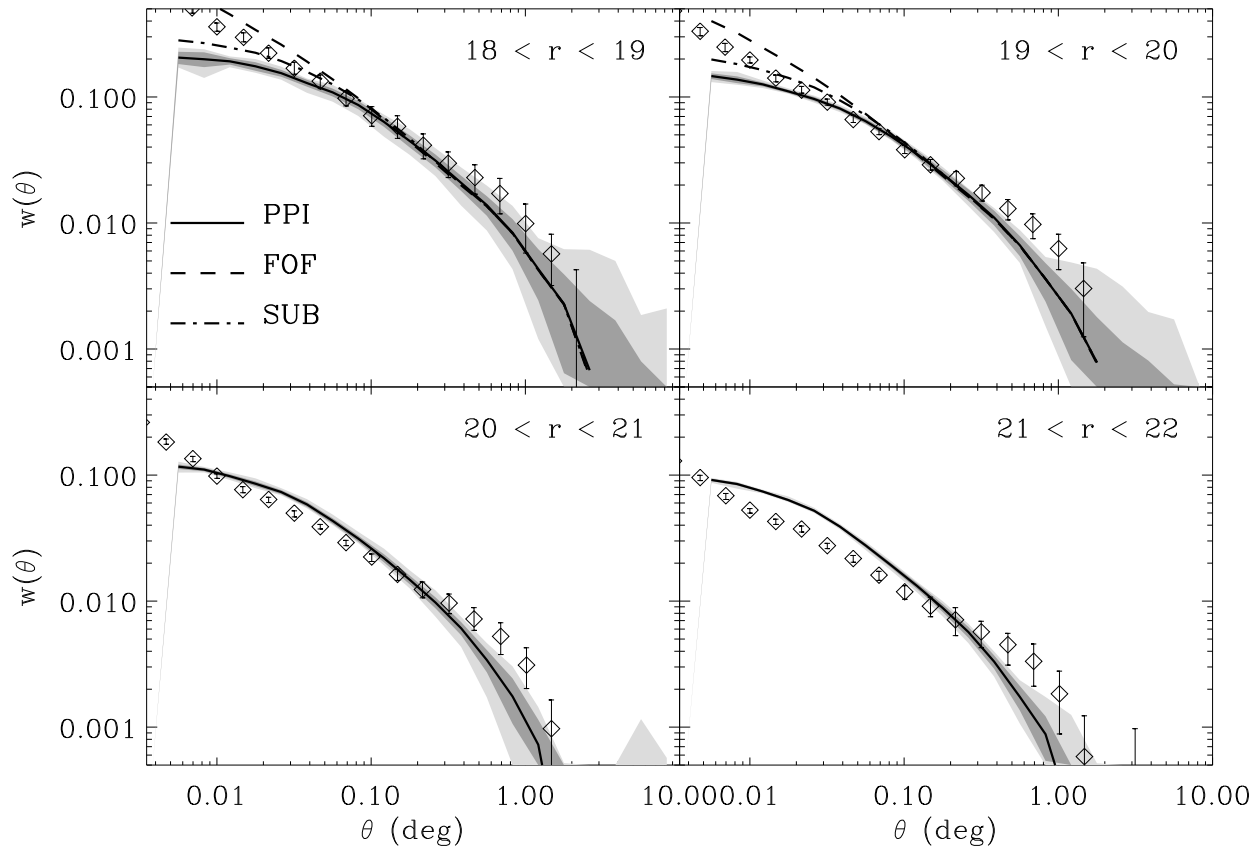,width=\textwidth}
\end{center}
\caption{Angular correlation function measured from our mock
catalogues for galaxies selected in four apparent magnitude bins. We
computed $w(\theta)$ for 20 independent observing cones of
2.5$\times$90 square degrees. The solid lines show the mean value, the
dark grey area shows the dispersion among the 20 cones, and the light
grey region shows the envelope of measures. In each panel, the
diamonds with error-bars show the SDSS ACF measured by
\citet{ConnollyEtal02}. \label{fig:w_std}}
\end{figure*}

In Fig. \ref{fig:w_std}, we show the mean ACF (solid line) and its
dispersion (dark grey region) from 20 mock catalogues built using
the PPI positioning scheme. The light grey areas show the envelopes of
the measurements. The diamonds with error bars are taken from 
\citet{ConnollyEtal02}. Now, some explanations might be useful to
understand how {\it good} the match actually is between our model and
the observations. 

{\it At large scales}, typically larger than $\sim \theta_b / 10$,
where $\theta_b$ is the angular size of our simulated volume at the
median redshift of a considered magnitude bin, finite volume effects
affect our estimates of the ACF (see Table \ref{tab:mocks} for
numerical values of $\theta_b$). This effect was discussed at length
in \citet{BlaizotEtal05} and is responsible for the (small)
under-estimate of the ACF at large scales. Interestingly, the finite
volume limit of our simulation neighbours that of the observations,
since the observed stripe is 2.5 degrees large. The disagreement of
our model with data at large scales is thus well understood in terms
of finite volume effects, hence it does not point to any failure in
the \galics{} model.

{\it At faint magnitudes}, especially in the two faintest apparent
magnitude bins, incompleteness -- due to mass resolution -- settles in
progressively, and is responsible for the increasing amplitude
over-estimate faint-wards. The three bottom lines of Table
\ref{tab:mocks} give us insight on this bias. As discussed in
Sec. \ref{sec:mocks}, faint-wards $r \sim 20$, one finds $M_r > I_{\tt
res}$, which means that the samples of galaxies selected in the two
faint magnitude bins are incomplete. This incompleteness is such that
the selected galaxies inhabit a population of haloes biased towards
high masses. Now, as massive haloes cluster more than low-mass ones,
increasing incompleteness implies an increasing positive bias in the
ACF. This is what happens at $r>20$. However, our results in the
apparent-magnitude range $[18;20]$ are robust, and the amplitude of
the ACF found for these samples is in good agreement with
observations.

{\it At small scales}, typically smaller than the angular size of a
group (i.e. $\sim \theta[1$ Mpc$]$ at $z_{\tt med}$) our predicted ACF
under-estimates the observed one. We will show in the next sub-section
that this bias can be attributed to an over-diluted distribution of
galaxies within haloes of DM. 

The discussion above shows that our results are in good agreement with
observations, in the rather restricted domain where our model and
catalogues are valid. We can nevertheless certainly improve the
situation, as we discuss in the following subsection.

\subsection{Exploring the small scale galaxy distribution} \label{sec:w_schemes}
The so-called ``halo occupation distribution'' (HOD) formalism has
proven to be quite helpful in terms of understanding the origin of the
clustering properties of galaxies. In the HOD framework, galaxy
clustering is the result of three ingredients : (i) the spatial
distribution of halos, (ii) the number of galaxies per halo, and (iii)
the distribution of galaxies within haloes. In the present study, the
distribution of haloes is drawn from a cosmological DM
simulation. Except for the very little difference between the
``concordance model'' and the cosmological parameters we assume, point
(i) is thus certainly the least questionable part of this work. The
number of galaxies that each halo harbours is a more difficult issue,
as it is the result of our complex semi-analytic
post-processing. Moreover, this quantity is very difficult to
constrain observationally \citep[see however][]{vdBoschYangMo03}. In a
paper in preparation, we compare the HOD obtained with \galics{} to
that obtained with a cosmological smoothed-particle hydrodynamics
(SPH) simulation, and find very good agreement. This, combined with
the numerous statistics that have been checked for our model
\citep{HattonEtal03,BlaizotEtal04,LanzoniEtal05} gives us confidence
in the fact that we predict the right number of galaxies per
halo. Then remains point (iii) only to explain the small-scale
discrepancy shown in the previous section between our model and
observations. One of the interesting results of the HOD formalism is
to decompose the correlation function into two terms. A term due to
pairs of galaxies located in different haloes (the 2-halo term)
dominates at large separations, and a term due to pairs of galaxies
populating the same halo (the 1-halo term) dominates the clustering
signal at small scales \citep[see e.g.][]{BerlindWeinberg02}. The
1-halo term is mainly due to galaxies that lie in groups or clusters,
and is sensitive to the way galaxies are spatially distributed within
these massive haloes.  Already from Fig. \ref{fig:positions}, one can
have a feeling of what is happening~: the distribution of galaxies
predicted by the PPI scheme within groups and clusters is less
concentrated than that obtained with the two other schemes (FOF and
SUB). This will naturally lead to an under-estimate of the 1-halo
term, and so to an under-estimate of the ACF at small scales.

In the top panels of Fig. \ref{fig:w_std}, we compare the ACFs
obtained with the three positioning schemes proposed in
Sec. \ref{sec:pos}. The solid (resp. dashed, dot-dashed) lines show
the mean PPI- (resp. FOF-, SUB-) ACF from the 20 mock catalogues
described in Sec. \ref{sec:mocks}.  This comparison tells us many
things. {\it First}, changing the distribution of galaxies within
haloes does indeed change the behaviour of the ACF at small
separations, although it leaves unchanged the ACF at large scales, as
expected.  {\it Second}, the FOF scheme yields an ACF which
over-estimates the observed ACF at small scales. If not a well known
result, this is at least a feature which is commonly found in the
literature \citep[e.g.][]{BensonEtal00, ScoccimarroEtal01,
BerlindEtal03, WeinbergEtal04, YangEtal04}. Although
\citet{YangEtal04} interpreted this feature as a hint that the
normalisation of the power-spectrum ($\sigma_8$) is over-estimated in
the concordance model, our analysis suggests another explanation which
simply relies on the distribution of galaxies within haloes. This
explanation is also supported by the work of \citet{KauffmannEtal99a}
who find a spatial correlation function in agreement with
observations. In their work, the positions of galaxies are obtained
following the most-bound particles of the haloes in which they were
formed. This is, in essence, similar to following sub-structures,
except that it allows to follow them below mass resolution, and to
bypass the expensive identification of sub-structures.
{\it Third}, the SUB scheme gives a result intermediate between FOF
and PPI, but still not in agreement with the data~: it yields a
depletion of the two-point correlation function at small separations,
similar to PPI. This was expected, as discussed in Sec.~\ref{sec:pos},
due to the fact that sub-structures are tidally stripped as they
spiral towards the centres of massive haloes.  As a result, the number
of pairs found at small separations with our SUB scheme is smaller
than what we would expect from the real galaxy distribution. This
effect is also increased by artificial phase-space heating due to
$N$-body relaxation. In reality, even if a sub-halo is tidally
stripped, its host galaxy still exists. However at variance with pure
dark matter, galaxies can experience non trivial collisions that would
expectingly reduce slightly their concentration in the centre of rich
haloes, which give a likely explanation for the fact that FOF
overestimates the ACF at small separations.

It is then hard to find a way to populate haloes with finite
resolution DM simulations only. Even following sub-haloes dynamically
as in \citet{SpringelEtal01} requires the use of a proxy when
sub-haloes dissolve : galaxies are then associated to the locally most
bound particle. It is however not clear how this proxy behaves once
sub-structures are disrupted -- although long relaxation times suggest
that the trajectories of once most-bound particles should be a good
approximation. 
In view of this effect, the
agreement found by \citet{KauffmannEtal99a} can be understood as the
result of an average between our FOF and SUB biasing schemes, confirming our
above statement that the SUB and FOF prescriptions narrowly bracket
the real solution.

To summarise the results, although the accuracy reached in this paper
cannot really help us to rigorously disentangle the SUB and FOF
schemes, our measurements confirm well known results of the
literature~: (i) sub-structures are non trivially biased tracers of
galaxies, and (ii) galaxies are distributed inside haloes very much
like dark matter, but in a slightly less concentrated way.
In the next section, we explore how this assertion resists the
additional constraints from higher-order clustering.

\section{Higher-order statistics} \label{sec:Sn}
Because gravity has long pulled structures harbouring galaxies away
from possible initial Gaussianity, the distribution of galaxies is not
fully characterised by the two-point correlation function
alone. Instead, higher-order correlations have become non-zero and
encapsulate the details of the small-scale non-linear galaxy
distribution.  It is thus very important to confront higher-order
predictions from our model to observational determinations. In this
section, we first explain the count-in-cells method that we used to
measure high-order clustering on SDSS DR1 and on mock
catalogues. Then, we briefly discuss the estimate made on SDSS
DR1. And finally, we compare results obtained with SDSS-DR1 and our
mocks in order to understand whether this new set of constraints can
help discriminate between our SUB and FOF schemes.

\subsection{The counts in cells method}

The probability distribution of counts in cells (CIC), $P_N(\theta)$,
is the probability that an angular cell of (linear) dimension $\theta$
contains $N$ galaxies.  The factorial moments of this distribution are
defined by $F_k \equiv \sum_N P_N (N)_k$, where $(N)_k =
N(N-1)..(N-k+1)$ is the $k$-th falling factorial of $N$.  The
factorial moments are closely related to the moments of the underlying
continuum random field (which is assumed Poisson-sampled by the
galaxies), $\rho = \avg{N}(1+\delta)$, through $\avg{(1+\delta)^k} =
F_k/\avg{N}^k$ \citep{SzapudiSzalay93}, where angle brackets in the
last relation denote an area average over cells of size $\theta$.  The
factorial moments therefore provide a convenient way to estimate the
angular connected moments, $S_p \equiv \langle\delta^p \rangle_c /
\langle \delta^2 \rangle^{p-1}$, where the subscript $c$ denotes the
connected contribution, and $\avg{\delta^p}_c$ denotes the area
average (over scale $\theta$) of the $p-$point angular correlation
function.  The moments $S_3$ (skewness) and $S_4$ (kurtosis) quantify
the lowest-order deviations of the angular distribution from a
Gaussian.

It is straightforward to calculate factorial moments from the
distribution of CIC, and one can then use the recursion relation of
\citet{SzapudiSzalay93} to obtain the $S_p$'s. This
technique is described in more complete detail in
\citet{SzapudiMeiksinNichol96} and \citet{SzapudiEtal01}. The most
delicate and time consuming component of estimating the cumulants
$S_p$'s is then the accurate measurement of CIC distribution.


As shown in \citet{SzapudiColombi96}, large-scale measurements are
dominated by edge effects which are impossible to correct for exactly
-- even when using massive over-sampling. This stems from the fact
that, due to finite cell size, galaxies near the edge of the survey
(or near a masked-out region) receive a smaller statistical weight
than galaxies away from any edge. This has devastating effect when
estimating CIC in galaxy surveys. Typically, across the whole SDSS
area, there are over 100 cut-out holes per sq. degree. Consequently, a
randomly placed cell of side $\sim0.1^\circ$ has a high probability of
intersecting a mask. Now, because traditional CIC techniques discard
such cells, they would not be able provide us with measurments on
scales larger than $\sim0.1$ degree. To remedy this situation, we
measure the CIC distribution using a new estimator by Colombi \&
Szapudi (2006, {\it in prep.}) and its implementation BMW-PN (for
Black-Magic-Weighted-PN). This estimator features a linear, massively
oversampling algorithm, and sports a new approximate edge correction
scheme.

The recipe implemented in BMW-PN gives approximately equal weight to
each galaxies during CIC estimation. While it was shown previously
that this is impossible under the most general circumstances, the
approximate scheme uses the fact that the CIC distribution is fairly
insensitive to cell shape (Szapudi 1998). This empirical fact can be
used for edge effect correction for the special case of estimating
galaxy CIC in the following way. The data are pixelized on a very fine
grid, which will give CIC for the smallest possible scale, the grid
step size. The same operation is performed for the masks. On these
pixelized data, one considers all the possible square cells of all
possible sizes, that are seen as ensembles of pixels. For each of
these cells, an effective size is given, corresponding to the valid
area it encompasses (overlapping pixelized masks are subtracted). Then
the center of mass of the valid part of the cell is calculated, and
one finds the pixel it falls into. With that procedure, a number of
cells of a given effective scale will fall onto this same
pixel. However, one is interested only in one cell, because one cell
per pixel is enough to extract all the available statistical
information at the chosen pixelization level. One thus selects the
cell which is the most compact one, or, in other words, the initial
square cell of smallest possible size before mask area
substraction. This way, one increases the effective area sampled by
the cells and the amount of available statistics. Moreover, due to the
fact that only one cell at most is allowed to contribute to a pixel, a
more even weight is given in practice to galaxies near the edge of the
catalog, which reduces edge effects. It is however not easy to
demonstrate that analytically: only practical experiments show that it
is indeed the case (see Colombi \& Szapudi 2006). This is why the
method is called ``Black-Magic Weighting'' (BMW).

A more detailed explanation of this estimator is given by Colombi \&
Szapudi (2006) who performed a series of tests based on simulated
galaxy surveys, and masks lifted from real galaxy surveys. They have
found that the method works with high precision. The control
parameter, a number between 0 and 1, determines the fraction of the
cell allowed to overlap with a mask. 0 corresponds to no overlap
(``classical'' CIC estimation), while larger numbers turn on the
BMW. It was found that even at 75\% allowed overlap, the systematic
errors introduced are negligible. For a margin of error, we have
allowed 50\% overlap in all the calculations presented below.

\subsection{The SDSS-DR1 data set} 

The first major SDSS data release \citep[DR1]{AbazajianEtal03} covers
2099 square degrees and contains over 53 million objects. In our
analysis, we include galaxies of the eight northern stripes 9-12,
34-37 and three southern ones 76, 82, 86 -- that is, all DR1 stripes
except the shortest ones, 42 and 43. While we use some data outside
the DR1, our area adds up to marginally smaller than the total area of
DR1\footnote{Note that DR4 is now publicly available. We however still
use DR1 because it is largely sufficient for the level of detail we
wish to reach, given the size of the simulation used.}. The galaxies
were split into four apparent magnitude bins that can be compared to
previous results of other surveys as well as the early SDSS
measurements by Szapudi et al 2002. The number of galaxies with
dereddened $\prime r$ model magnitudes between 18--19, 19--20, 20--21
and 21--22 are 732,216, 2,047,766, 5,455,559 and 10,890,300,
respectively. That is altogether more than 19.1 million galaxies,
which is an order of magnitude more than the largest higher order
statistical study to date.

The database also holds the relevant information about areas on the
sky that are to be censored in any type of statistical studies of
spatial distribution of galaxies. Bright stars, satelites, airplanes
and bad seeing account for approximately 12\% loss in the area of DR1.
These masked regions on the sky are extremely hard to deal with in CIC
measurements, as explained below.

Since we have measured CIC in 11 virtually independent SDSS stripes,
we were able to estimate the variance in a fairly robust fashion, by
taking the unbiased dispersion over the 11 stripes and dividing
the corresponding error by a factor $\sqrt{11}$ (see e.g.
Colombi, Szapudi \& Szalay 1998). All
errorbars have been determined this way. For the mock catalogs, the
errors are determined from the dispersion obtained from 20 random stripes without
further renormalization: the error estimated this way assumes only one stripe.
Indeed, the simulation used to generate the mock catalogs is too small
to have fair estimate of the errors on a full 11 stripes catalog.

We performed a series of measurements on the DR1 data set, using the
CIC method. We extracted $S_3$ to $S_5$ from the CIC statistics. All
the measurements were carried out with masks corresponding to seeing
limit $FWHM=1.7,1.8,1.9,2.0$ arcsec. 

\begin{figure*}
\begin{center}
\psfig{file=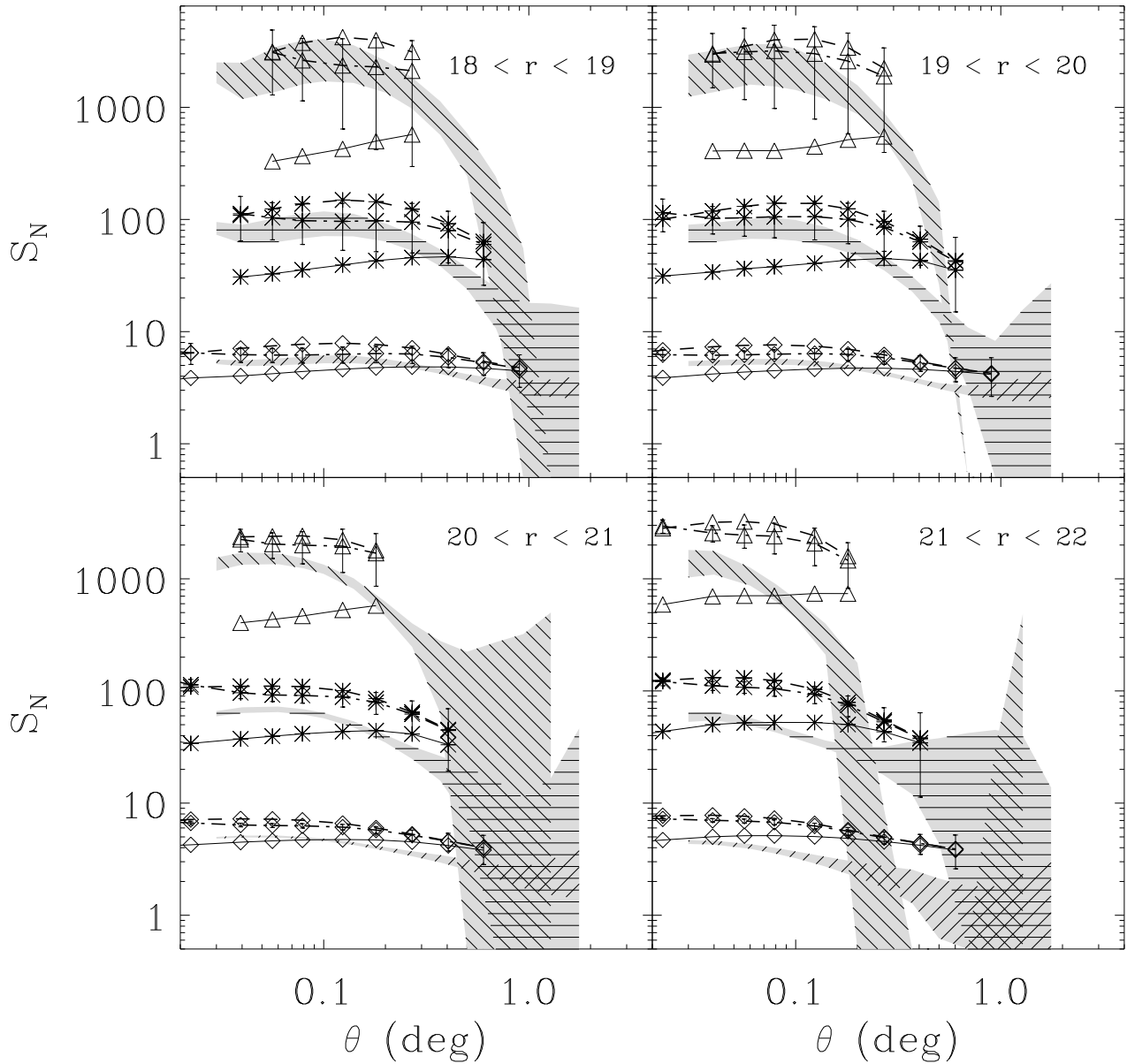,width=\textwidth}
\end{center}
\caption{Third (resp. fourth, fifth) order cumulants measured in 20
mock catalogues are shown as diamonds (resp. stars, triangles). Solid
(resp. dashed, dot-dashed) lines connect symbols corresponding to the
PPI (resp. FOF, SUB) scheme. The error bars associated with symbols
show the dispersion around the average estimate. The four panels
correspond to different apparent magnitude selections, as in
Fig. \ref{fig:w_std}. The shaded regions show the locus of SDSS
measurements.\label{fig:Sn}}
\end{figure*}

We have found that seeing has
only minor effects for the most part, therefore we present
measurements for $FWHM=1.7$ only (shaded areas in Fig. \ref{fig:Sn}).

\subsection{Mocks vs. DR1} \label{sec:sn_schemes}

In Fig. \ref{fig:Sn}, diamonds (resp. stars, triangles) show the mean
values of $S_3$ (resp. $S_4$, $S_5$), obtained from the 20 mock
catalogues. The symbols connected with continuous (resp. dashed,
dot-dashed) lines correspond to the PPI (resp. FOF, SUB) schemes for
positioning galaxies within haloes.  The dispersion of $S_n$ estimates
are shown with error bars for the SUB case only, for the sake of
clarity.  Measurements from the SDSS DR1 are shown with the shaded and
hatched areas. These were obtained with the 1.7 arcseconds seeing masks.
Note that errors computed for the predictions are larger than for the
data. 
As already mentionned above, 
the reason for that is that for the first case, the errors are
obtained from the dispersion over the 20 simulated stripes, while the
errors in the second case take into account the fact that there are 11
stripes in the DR1 survey, hence corresponding to errors $\sim
\sqrt{10} \sim 3$ times smaller. Note that some data points are not
shown from the model at large and small scales. Points were removed
when the associated error bars became too large.

As for the 2-point angular correlation, model results in the two
faintest magnitude bins are strongly affected by incompleteness. This
again leads to an over-estimate of the $S_n$ coefficients, increasing
with apparent magnitude. Similarly, large scales are affected by
finite volume effects. The poor agreement between the models and the
observations at large scales, even on the upper panels of
Fig.~\ref{fig:Sn} can thus certainly be blamed on finite volume/edge
effects as mentioned earlier. It remains quite acceptable given the
error bars.  The fact that cumulants from the three different
positioning schemes converge only at rather large scales -- larger
than for the 2-point correlation function -- is just because these
quantities are cell averages, i.e. integrated from $0$ to $\theta$.
At brighter magnitudes ($r<20$), the model's predictions are robust,
as discussed for the ACF. Fig. \ref{fig:Sn} then tells us the
following.
\begin{description}
\item[(i)] The FOF and SUB schemes show a similar good agreement with
observations and are almost indistinguishable from each other given
the level of uncertainty on the measurements. In principle, one might
expect significant differences between FOF and SUB at small scales, as
found for the two-point correlation function, but the effect seems be
of the same order of magnitude on $\avg{\delta^N}_c$ and $\avg{\delta^2}^{N-1}$
and thus disappears in the normalisation.
\item[(ii)] We recover in the two upper panels of Fig. \ref{fig:Sn}
the fact found in the previous section, that the PPI scheme leads to
an expected strong under-estimate of the observed $S_n$
coefficients. The effect is the strongest at small scales, while PPI
converges to FOF and SUB at large scales, when data points are
available.
\item[(iii) ] Finally, the agreement of \galics{} with the DR1
estimates is quite a succes, provided that modelled galaxies are
distributed as sub-structures or DM within haloes.
\end{description}

\section{Conclusions} \label{sec:conclu}

In this paper, we have used a novel technique for constructing mock
SDSS-like observations from the predictions of a hybrid model of
galaxy formation \citep{HattonEtal03,BlaizotEtal05}. Although mock
observations have been made in the past from hybrid models of galaxy
formation \citep[e.g.][]{DiaferioEtal99,MathisEtal02}, we emphasize
that our method is general and can readily be used to reproduce {\it
any} type of extra-galactic survey.  We have used these mock
observations to carry out a detailed comparison of the clustering
properties of galaxies observed in the SDSS to those predicted by a
state-of-the-art implementation of the hierarchical galaxy formation
scenario. We have carefully investigated the limitations of our model,
which are mostly due to mass resolution and finite volume of the DM
simulation. Mass resolution directly translates into incompleteness at
faint apparent magnitudes, such that the selected galaxies inhabit
haloes biased towards high masses. This in turn leads to an increasing
over-estimate of the clustering statistics faintwards. Our predictions
are robust, though, at magnitudes brighter than $r \sim 20$. The
finite volume of the simulation introduces a negative bias in
clustering statistics at large scales, well known as the ``integral
constrain'' problem. We have shown in \citet{BlaizotEtal05} how this
affects angular correlation function estimates from mock catalogues,
and can thus safely define a safe validity scale range for our
predictions, which typically extends up to a tenth of the apparent
size of the simulated volume at the median redshift of the selected
sample.  Within the rather restricted domain where the model
predictions are valid, we find a good agreement with the observed
angular two-point correlation function.

At small scales -- typically $<1 h^{-1}$Mpc -- our standard PPI
positioning scheme is found to under-estimate the 2-point correlation
function. This can be explained by the fact that this modelling of
galaxy positions within haloes yields too diluted a distribution of
galaxies within groups and clusters. We thus investigated the impact
of changing the spatial distribution of galaxies within haloes on the
ACF and found that observations can be explained if galaxies have a
distribution somewhere between that of DM particles and DM
sub-structures, as suggested by the early results of
\citet{KauffmannEtal99a}.

Moving to higher-order statistics, we can robustly rule out the PPI
scheme. The uncertainty on the measurments do not allow to
discriminate between the FOF and SUB schemes. This work shows that
modelling the distribution of galaxies within massive haloes is a
difficult task. In particular, the instantaneous view of the DM
distribution is not enough (yet) to populate haloes, because (i) the
positions of galaxies are the result of an evolutionary process and
(ii) the sub-structures that harbour them dissolve if prohibitively
high resolution is not used. A straight-forward biasing scheme based
either on DM particles or on DM sub-structures is thus, as already
found in the literature, a poor proxy for positioning galaxies, and we
show that it leads to an over-estimate (respectively an
under-estimate) of the clustering signal at small
separations. Existing attempts to follow the dynamics of galaxies
within DM-only simulations still suffer from resolution effects
\citep{SpringelEtal01}. As a result, most galaxies in the core regions
of massive haloes are attached to particles (once most-bound) rather
than sub-structures. This makes it necessary for HOD models to
incorporate in some way the evolution of the sub-haloes, e.g. by
keeping track of once most-bound particles, just like in SAMs.

Most of the limitations of the present work are due to the rather
small range where our model's predictions are robust, which is in turn
mainly due to the properties of the DM simulation we use. Namely, mass
resolution and finite volume effects do not allow us to make full use
of the wealth of data obtained by the SDSS. One obvious way to improve
the situation is to use bigger simulations. In this prospect, the
so-called ``millennium simulation'' from the Virgo
consortium\footnote{{\tt http://www.virgo.dur.ac.uk/}} will
undoubtedly help us progress on the interpretation of the clustering
properties of galaxies in the nearby Universe. The mass resolution of
this simulation is about 10 times better than that of the simulation
used in this work. This should allow us to make better use of the
observations in the apparent-magnitude range $20<r<22$. And the volume
of the millennium simulation is 125 times larger, which should allow
(i) better estimates of cosmic variance, and (ii) robust
characterisation of the large scale distribution of galaxies
(aleviating finite volume effects). The sheer statistics from this
simulation should also allow us to carry out a more subtle study of
the dependence of galaxy clustering on various galaxy properties
(e.g. luminosities, colors, age, morphological types, etc.), thereby
allowing to set constraints on the baryonic physics of galaxy
formation.

\section*{Acknowledgements}
The authors thank D. H. Weinberg, S. D. M. White and V. Springel for
many enlightening discussions.  The N-body simulation used in this
work was run on the Cray T3E at the IDRIS super-computing
facility. This work was performed in the framework of the HORIZON
project.


\begin{thebibliography}{}
\bibitem[\protect\citeauthoryear{{Abazajian}, {Adelman-McCarthy}, {Ag{\"u}eros}
  et~al.,}{{Abazajian} et~al.}{2003}]{AbazajianEtal03}
{Abazajian} K.,  {Adelman-McCarthy} J.~K.,  {Ag{\"u}eros} M.~A.,    et~al.,
  2003, \aj, 126, 2081

\bibitem[\protect\citeauthoryear{{Aubert}, {Pichon} \& {Colombi}}{{Aubert}
  et~al.}{2004}]{AubertPichonColombi04}
{Aubert} D.,  {Pichon} C.,    {Colombi} S.,  2004, \mnras, 352, 376

\bibitem[\protect\citeauthoryear{{Bardeen}, {Bond}, {Kaiser} \&
  {Szalay}}{{Bardeen} et~al.}{1986}]{BardeenEtal86}
{Bardeen} J.~M.,  {Bond} J.~R.,  {Kaiser} N.,    {Szalay} A.~S.,  1986, \apj,
  304, 15

\bibitem[\protect\citeauthoryear{{Benson}, {Cole}, {Frenk}, {Baugh} \&
  {Lacey}}{{Benson} et~al.}{2000}]{BensonEtal00}
{Benson} A.~J.,  {Cole} S.,  {Frenk} C.~S.,  {Baugh} C.~M.,    {Lacey} C.~G.,
  2000, \mnras, 311, 793

\bibitem[\protect\citeauthoryear{{Benson}, {Frenk}, {Baugh}, {Cole} \&
  {Lacey}}{{Benson} et~al.}{2001}]{BensonEtal01}
{Benson} A.~J.,  {Frenk} C.~S.,  {Baugh} C.~M.,  {Cole} S.,    {Lacey} C.~G.,
  2001, \mnras, 327, 1041

\bibitem[\protect\citeauthoryear{{Berlind} \& {Weinberg}}{{Berlind} \&
  {Weinberg}}{2002}]{BerlindWeinberg02}
{Berlind} A.~A.,  {Weinberg} D.~H.,  2002, \apj, 575, 587

\bibitem[\protect\citeauthoryear{{Berlind}, {Weinberg}, {Benson}, {Baugh},
  {Cole}, {Dav{\' e}}, {Frenk}, {Jenkins}, {Katz} \& {Lacey}}{{Berlind}
  et~al.}{2003}]{BerlindEtal03}
{Berlind} A.~A.,  {Weinberg} D.~H.,  {Benson} A.~J.,  {Baugh} C.~M.,  {Cole}
  S.,  {Dav{\' e}} R.,  {Frenk} C.~S.,  {Jenkins} A.,  {Katz} N.,    {Lacey}
  C.~G.,  2003, \apj, 593, 1

\bibitem[\protect\citeauthoryear{{Blaizot}, {Guiderdoni}, {Devriendt},
  {Bouchet}, {Hatton} \& {Stoehr}}{{Blaizot} et~al.}{2004}]{BlaizotEtal04}
{Blaizot} J.,  {Guiderdoni} B.,  {Devriendt} J.~E.~G.,  {Bouchet} F.~R.,
  {Hatton} S.~J.,    {Stoehr} F.,  2004, \mnras, 352, 571

\bibitem[\protect\citeauthoryear{{Blaizot}, {Wadadekar}, {Guiderdoni},
  {Colombi}, {Bertin}, {Bouchet}, {Devriendt} \& {Hatton}}{{Blaizot}
  et~al.}{2005}]{BlaizotEtal05}
{Blaizot} J.,  {Wadadekar} Y.,  {Guiderdoni} B.,  {Colombi} S.~T.,  {Bertin}
  E.,  {Bouchet} F.~R.,  {Devriendt} J.~E.~G.,    {Hatton} S.,  2005, \mnras,
  360, 159

\bibitem[\protect\citeauthoryear{{Cen} \& {Ostriker}}{{Cen} \&
  {Ostriker}}{2000}]{CenOstriker00}
{Cen} R.,  {Ostriker} J.~P.,  2000, \apj, 538, 83

\bibitem[\protect\citeauthoryear{{Colombi} \& {Szapudi}}{{Colombi} \&
  {Szapudi}}{2006}]{ColSza06}
{Colombi} S.,  {Szapudi} I.,  2006, in preparation

\bibitem[\protect\citeauthoryear{{Colombi}, {Szapudi} \& {Szalay}}{{Colombi}
  {Szapudi} \& {Szalay}}{1998}]{ColSzaSza98}
{Colombi} S.,  {Szapudi} I.,  {Szalay} A.~S., 1998, \mnras, 296, 253

\bibitem[\protect\citeauthoryear{{Connolly}, {Scranton}, {Johnston}
  et~al.,}{{Connolly} et~al.}{2002}]{ConnollyEtal02}
{Connolly} A.~J.,  {Scranton} R.,  {Johnston} D.,    et~al., 2002, \apj, 579,
  42

\bibitem[\protect\citeauthoryear{{Davis}, {Efstathiou}, {Frenk} \&
  {White}}{{Davis} et~al.}{1985}]{DavisEtal85}
{Davis} M.,  {Efstathiou} G.,  {Frenk} C.~S.,    {White} S.~D.~M.,  1985, \apj,
  292, 371

\bibitem[\protect\citeauthoryear{{De Lucia}, {Kauffmann} \& {White}}{{De Lucia}
  et~al.}{2004}]{DeLuciaKauffmannWhite04}
{De Lucia} G.,  {Kauffmann} G.,    {White} S.~D.~M.,  2004, \mnras, 349, 1101

\bibitem[\protect\citeauthoryear{{Devriendt}, {Blaizot}, {Guiderdoni},
  {Hatton}, {Ninin} \& {Bouchet}}{{Devriendt} et~al.}{2005}]{DevriendtEtal05}
{Devriendt} J.~E.~G.,  {Blaizot} J.,  {Guiderdoni} B.,  {Hatton} S.,  {Ninin}
  S.,    {Bouchet} F.~R.,  2005, in preparation

\bibitem[\protect\citeauthoryear{{Devriendt}, {Guiderdoni} \&
  {Sadat}}{{Devriendt} et~al.}{1999}]{DevriendtGuiderdoniSadat99}
{Devriendt} J.~E.~G.,  {Guiderdoni} B.,    {Sadat} R.,  1999, \aap, 350, 381

\bibitem[\protect\citeauthoryear{{Diaferio}, {Kauffmann}, {Balogh}, {White},
  {Schade} \& {Ellingson}}{{Diaferio} et~al.}{2001}]{DiaferioEtal01}
{Diaferio} A.,  {Kauffmann} G.,  {Balogh} M.~L.,  {White} S.~D.~M.,  {Schade}
  D.,    {Ellingson} E.,  2001, \mnras, 323, 999

\bibitem[\protect\citeauthoryear{{Diaferio}, {Kauffmann}, {Colberg} \&
  {White}}{{Diaferio} et~al.}{1999}]{DiaferioEtal99}
{Diaferio} A.,  {Kauffmann} G.,  {Colberg} J.~M.,    {White} S.~D.~M.,  1999,
  \mnras, 307, 537

\bibitem[\protect\citeauthoryear{{Diemand}, {Moore} \& {Stadel}}{{Diemand}
  et~al.}{2004}]{DiemandMooreStadel04}
{Diemand} J.,  {Moore} B.,    {Stadel} J.,  2004, \mnras, 352, 535

\bibitem[\protect\citeauthoryear{{Eisenstein} \& {Hut}}{{Eisenstein} \&
  {Hut}}{1998}]{EisensteinHut98}
{Eisenstein} D.~J.,  {Hut} P.,  1998, \apj, 498, 137

\bibitem[\protect\citeauthoryear{{Eke}, {Cole} \& {Frenk}}{{Eke}
  et~al.}{1996}]{EkeColeFrenk96}
{Eke} V.~R.,  {Cole} S.,    {Frenk} C.~S.,  1996, \mnras, 282, 263

\bibitem[\protect\citeauthoryear{{Gao}, {De Lucia}, {White} \& {Jenkins}}{{Gao}
  et~al.}{2004}]{GaoEtal04}
{Gao} L.,  {De Lucia} G.,  {White} S.~D.~M.,    {Jenkins} A.,  2004, \mnras,
  352, L1

\bibitem[\protect\citeauthoryear{{Hatton}, {Devriendt}, {Ninin}, {Bouchet},
  {Guiderdoni} \& {Vibert}}{{Hatton} et~al.}{2003}]{HattonEtal03}
{Hatton} S.,  {Devriendt} J.~E.~G.,  {Ninin} S.,  {Bouchet} F.~R.,
  {Guiderdoni} B.,    {Vibert} D.,  2003, \mnras, 343, 75

\bibitem[\protect\citeauthoryear{{Helly}, {Cole}, {Frenk}, {Baugh}, {Benson} \&
  {Lacey}}{{Helly} et~al.}{2003}]{HellyEtal03a}
{Helly} J.~C.,  {Cole} S.,  {Frenk} C.~S.,  {Baugh} C.~M.,  {Benson} A.,
  {Lacey} C.,  2003, \mnras, 338, 903

\bibitem[\protect\citeauthoryear{{Jing}, {Mo} \& {Boerner}}{{Jing}
  et~al.}{1998}]{JingMoBoerner98}
{Jing} Y.~P.,  {Mo} H.~J.,    {Boerner} G.,  1998, \apj, 494, 1

\bibitem[\protect\citeauthoryear{{Kauffmann}, {Colberg}, {Diaferio} \&
  {White}}{{Kauffmann} et~al.}{1999a}]{KauffmannEtal99a}
{Kauffmann} G.,  {Colberg} J.~M.,  {Diaferio} A.,    {White} S.~D.~M.,  1999a,
  \mnras, 303, 188

\bibitem[\protect\citeauthoryear{{Kauffmann}, {Colberg}, {Diaferio} \&
  {White}}{{Kauffmann} et~al.}{1999b}]{KauffmannEtal99b}
{Kauffmann} G.,  {Colberg} J.~M.,  {Diaferio} A.,    {White} S.~D.~M.,  1999b,
  \mnras, 307, 529

\bibitem[\protect\citeauthoryear{{Kauffmann}, {Nusser} \&
  {Steinmetz}}{{Kauffmann} et~al.}{1997}]{KauffmannNusserSteinmetz97}
{Kauffmann} G.,  {Nusser} A.,    {Steinmetz} M.,  1997, \mnras, 286, 795

\bibitem[\protect\citeauthoryear{{Landy} \& {Szalay}}{{Landy} \&
  {Szalay}}{1993}]{LandySzalay93}
{Landy} S.~D.,  {Szalay} A.~S.,  1993, \apj, 412, 64

\bibitem[\protect\citeauthoryear{{Lanzoni}, {Guiderdoni}, {Mamon}, {Devriendt}
  \& {Hatton}}{{Lanzoni} et~al.}{2005}]{LanzoniEtal05}
{Lanzoni} B.,  {Guiderdoni} B.,  {Mamon} G.~A.,  {Devriendt} J.~E.~G.,
  {Hatton} S.,  2005, in preparation

\bibitem[\protect\citeauthoryear{{Mathis}, {Lemson}, {Springel}, {Kauffmann},
  {White}, {Eldar} \& {Dekel}}{{Mathis} et~al.}{2002}]{MathisEtal02}
{Mathis} H.,  {Lemson} G.,  {Springel} V.,  {Kauffmann} G.,  {White} S.~D.~M.,
  {Eldar} A.,    {Dekel} A.,  2002, \mnras, 333, 739

\bibitem[\protect\citeauthoryear{{Mathis} \& {White}}{{Mathis} \&
  {White}}{2002}]{MathisWhite02}
{Mathis} H.,  {White} S.~D.~M.,  2002, \mnras, 337, 1193

\bibitem[\protect\citeauthoryear{{Nagai} \& {Kravtsov}}{{Nagai} \&
  {Kravtsov}}{2005}]{NagaiKravtsov05}
{Nagai} D.,  {Kravtsov} A.~V.,  2005, \apj, 618, 557

\bibitem[\protect\citeauthoryear{{Ninin}}{{Ninin}}{1999}]{Ninin99}
{Ninin} S.,  1999, PhD thesis, Universit\'e Paris 11

\bibitem[\protect\citeauthoryear{{Peacock} \& {Smith}}{{Peacock} \&
  {Smith}}{2000}]{PeacockSmith00}
{Peacock} J.~A.,  {Smith} R.~E.,  2000, \mnras, 318, 1144

\bibitem[\protect\citeauthoryear{{Pearce}, {Jenkins}, {Frenk}, {Colberg},
  {White}, {Thomas}, {Couchman}, {Peacock}, {Efstathiou} \& {The Virgo
  Consortium}}{{Pearce} et~al.}{1999}]{PearceEtal99}
{Pearce} F.~R.,  {Jenkins} A.,  {Frenk} C.~S.,  {Colberg} J.~M.,  {White}
  S.~D.~M.,  {Thomas} P.~A.,  {Couchman} H.~M.~P.,  {Peacock} J.~A.,
  {Efstathiou} G.,    {The Virgo Consortium} 1999, \apjl, 521, L99

\bibitem[\protect\citeauthoryear{{Pearce}, {Jenkins}, {Frenk}, {White},
  {Thomas}, {Couchman}, {Peacock} \& {Efstathiou}}{{Pearce}
  et~al.}{2001}]{PearceEtal01}
{Pearce} F.~R.,  {Jenkins} A.,  {Frenk} C.~S.,  {White} S.~D.~M.,  {Thomas}
  P.~A.,  {Couchman} H.~M.~P.,  {Peacock} J.~A.,    {Efstathiou} G.,  2001,
  \mnras, 326, 649

\bibitem[\protect\citeauthoryear{{Scoccimarro} \& {Sheth}}{{Scoccimarro} \&
  {Sheth}}{2002}]{ScoccimarroSheth02}
{Scoccimarro} R.,  {Sheth} R.~K.,  2002, \mnras, 329, 629

\bibitem[\protect\citeauthoryear{{Scoccimarro}, {Sheth}, {Hui} \&
  {Jain}}{{Scoccimarro} et~al.}{2001}]{ScoccimarroEtal01}
{Scoccimarro} R.,  {Sheth} R.~K.,  {Hui} L.,    {Jain} B.,  2001, \apj, 546, 20

\bibitem[\protect\citeauthoryear{{Scranton}, {Johnston}, {Dodelson}
  et~al.,}{{Scranton} et~al.}{2002}]{ScrantonEtal02}
{Scranton} R.,  {Johnston} D.,  {Dodelson} S.,    et~al., 2002, \apj, 579, 48

\bibitem[\protect\citeauthoryear{{Somerville}, {Lemson}, {Sigad}, {Dekel},
  {Kauffmann} \& {White}}{{Somerville} et~al.}{2001}]{SomervilleEtal01}
{Somerville} R.~S.,  {Lemson} G.,  {Sigad} Y.,  {Dekel} A.,  {Kauffmann} G.,
  {White} S.~D.~M.,  2001, \mnras, 320, 289

\bibitem[\protect\citeauthoryear{{Springel}, {White}, {Tormen} \&
  {Kauffmann}}{{Springel} et~al.}{2001}]{SpringelEtal01}
{Springel} V.,  {White} S.~D.~M.,  {Tormen} G.,    {Kauffmann} G.,  2001,
  \mnras, 328, 726

\bibitem[\protect\citeauthoryear{{Szapudi} \& {Colombi}}{{Szapudi} \&
  {Colombi}}{1996}]{SzapudiColombi96}
{Szapudi} I.,  {Colombi} S.,  1996, \apj, 470, 131

\bibitem[\protect\citeauthoryear{{Szapudi}, {Meiksin} \& {Nichol}}{{Szapudi}
  et~al.}{1996}]{SzapudiMeiksinNichol96}
{Szapudi} I.,  {Meiksin} A.,    {Nichol} R.~C.,  1996, \apj, 473, 15

\bibitem[\protect\citeauthoryear{{Szapudi}, {Postman}, {Lauer} \&
  {Oegerle}}{{Szapudi} et~al.}{2001}]{SzapudiEtal01}
{Szapudi} I.,  {Postman} M.,  {Lauer} T.~R.,    {Oegerle} W.,  2001, \apj, 548,
  114

\bibitem[\protect\citeauthoryear{{Szapudi}, {Prunet} \& {Colombi}}{{Szapudi}
  et~al.}{2001}]{SzapudiPrunetColombi01}
{Szapudi} I.,  {Prunet} S.,    {Colombi} S.,  2001, \apjl, 561, L11

\bibitem[\protect\citeauthoryear{{Szapudi} \& {Szalay}}{{Szapudi} \&
  {Szalay}}{1993}]{SzapudiSzalay93}
{Szapudi} I.,  {Szalay} A.~S.,  1993, \apj, 408, 43

\bibitem[\protect\citeauthoryear{Szapudi}{1998}]{1998ApJ...497...16S} 
Szapudi I., 1998, ApJ, 497, 16 

\bibitem[\protect\citeauthoryear{{van den Bosch}, {Yang} \& {Mo}}{{van den
  Bosch} et~al.}{2003}]{vdBoschYangMo03}
{van den Bosch} F.~C.,  {Yang} X.,    {Mo} H.~J.,  2003, \mnras, 340, 771

\bibitem[\protect\citeauthoryear{{Weinberg}, {Dav{\'e}}, {Katz} \&
  {Hernquist}}{{Weinberg} et~al.}{2004}]{WeinbergEtal04}
{Weinberg} D.~H.,  {Dav{\'e}} R.,  {Katz} N.,    {Hernquist} L.,  2004, \apj,
  601, 1

\bibitem[\protect\citeauthoryear{{Yang}, {Mo}, {Jing}, {van den Bosch} \&
  {Chu}}{{Yang} et~al.}{2004}]{YangEtal04}
{Yang} X.,  {Mo} H.~J.,  {Jing} Y.~P.,  {van den Bosch} F.~C.,    {Chu} Y.,
  2004, \mnras, 350, 1153

\bibitem[\protect\citeauthoryear{{Yasuda}, {Fukugita}, {Narayanan}
  et~al.,}{{Yasuda} et~al.}{2001}]{YasudaEtal01}
{Yasuda} N.,  {Fukugita} M.,  {Narayanan} V.~K.,    et~al., 2001, \aj, 122,
  1104

\bibitem[\protect\citeauthoryear{{York} et~al.,}{{York}
  et~al.}{2000}]{YorkEtal00}
{York} D.~G.,  et~al., 2000, \aj, 120, 1579

\bibitem[\protect\citeauthoryear{{Yoshikawa}, {Taruya}, {Jing} \&
  {Suto}}{{Yoshikawa} et~al.}{2001}]{YoshikawaEtal01}
{Yoshikawa} K.,  {Taruya} A.,  {Jing} Y.~P.,    {Suto} Y.,  2001, \apj, 558,
  520
\end{thebibliography}

\appendix
\section{Substructures} \label{sec:HOP}

\begin{figure*}
\begin{tabular}{cc}
\psfig{file=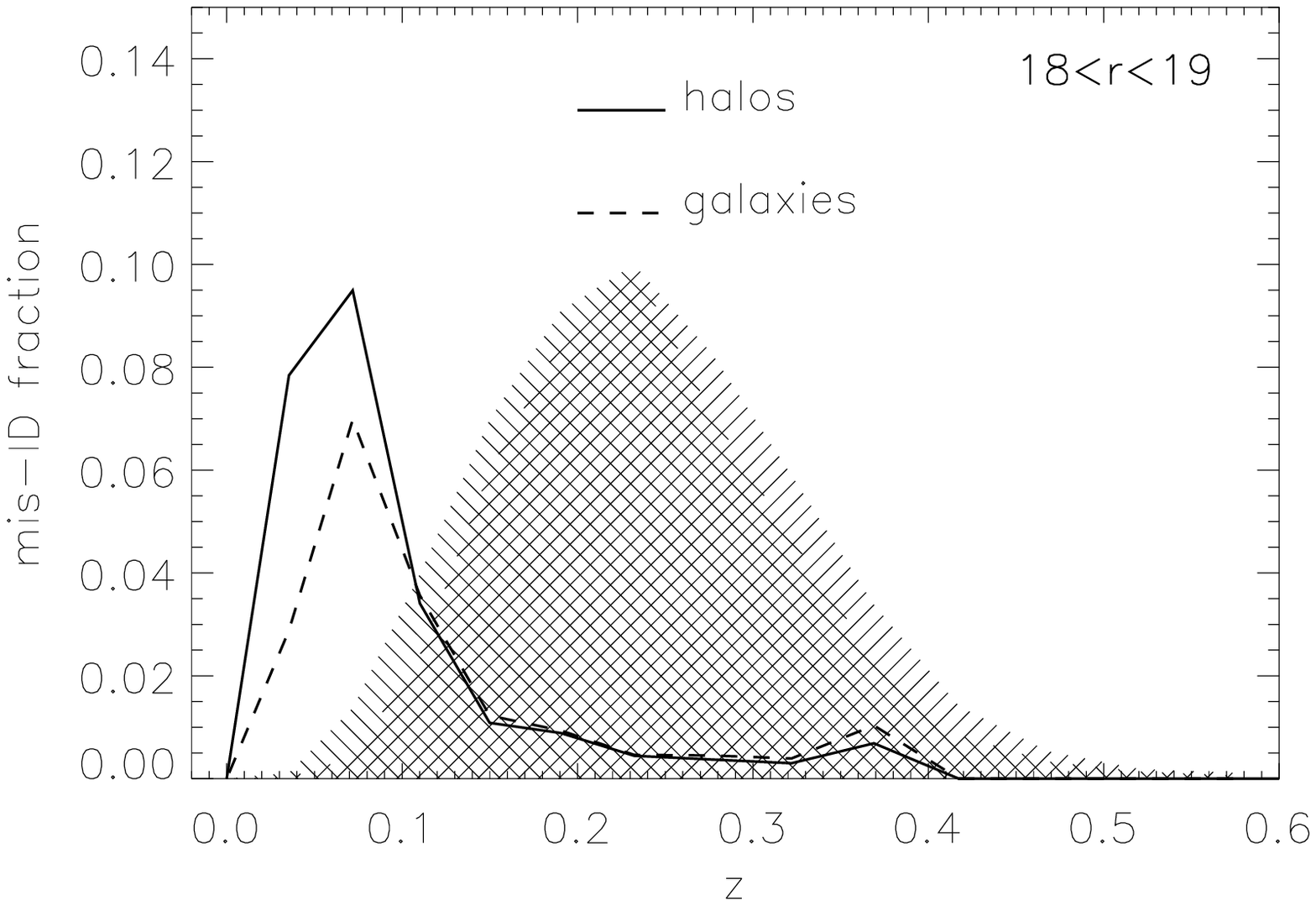,width=0.48 \textwidth} & 
\psfig{file=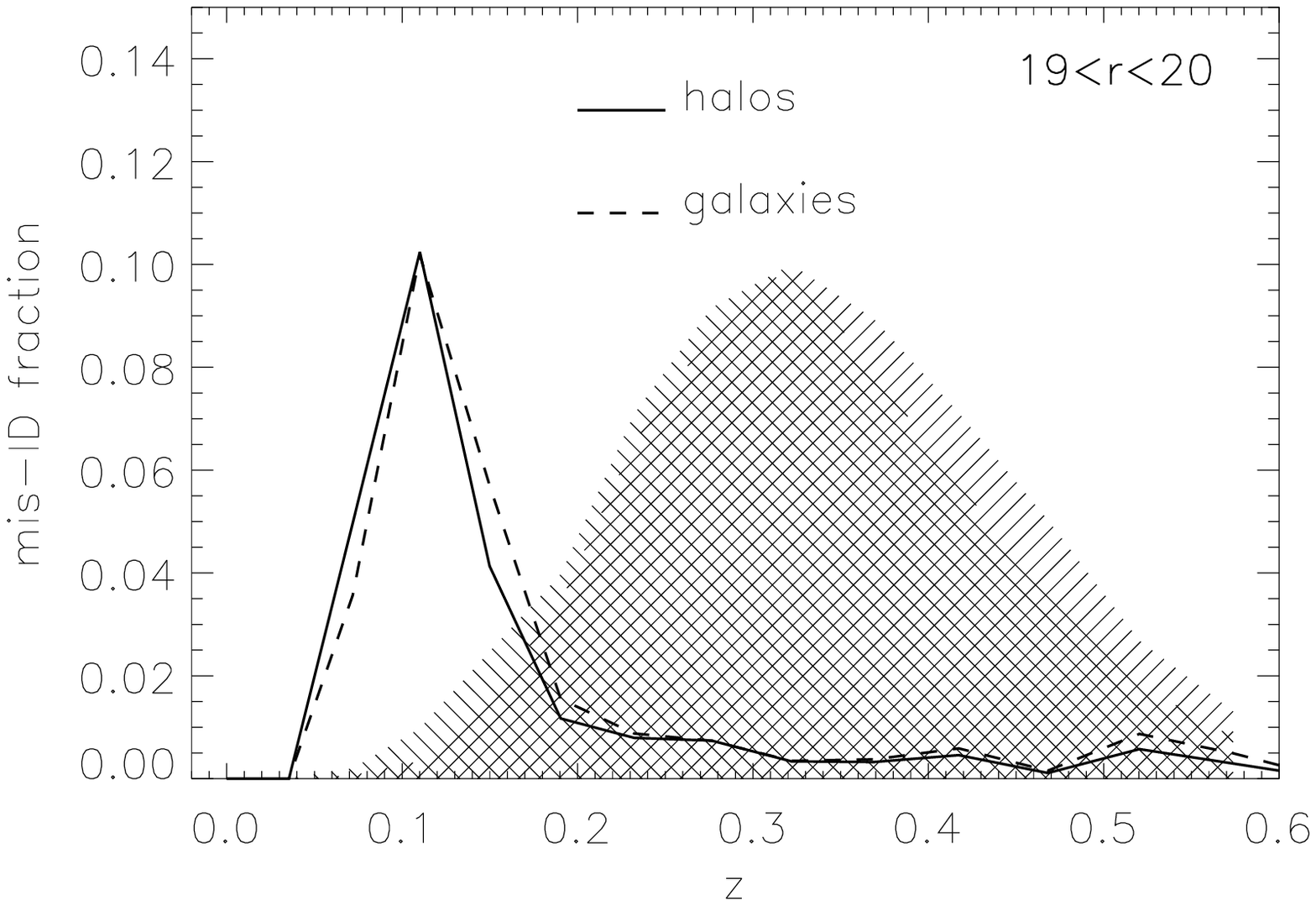,width=0.48 \textwidth} 
\end{tabular}
\caption{Dashed lines show the fraction of selected galaxies which are
not associated with a sub-structure. Solid lines show the fraction of
haloes containing more detected galaxies than sub-structures. The
hatched area shows the arbitrarily scaled red-shift distribution of
galaxies (down-right hatches) and haloes (up-right hatches) in the
catalogue. Less than 1\% of galaxies are not associated with
sub-structures in our catalogues, at $r<20$. NB : in the
right-hand-side panel, the truncation of the red-shift distributions
is only an artifact of the plotting routine.\label{fig:misidfracs}}
\end{figure*}

\subsection{Identification}

We identified sub-structures in our DM simulation using the \ahop{}
code \citep[][ appendix B.]{AubertPichonColombi04}, which is an
extension of the HOP halo finder \citep{EisensteinHut98}. This
algorithm exploits basic principles of the Morse theory to extract a
tree of structures (the halos) and sub-structures from a distribution
of DM particles. It proceeds in four steps, which are the following.
\begin{itemize}
\item[(i)] One needs to estimate the local density associated to each
DM particle using smoothed-particle hydrodynamics (SPH) interpolation
over $N_{\rm SPH}$ neighbours. Here, we take $N_{\rm SPH}=20$ in order
to match the FOF halo population, as explained below. During this
process, one should store the $N_{\rm HOP}$ nearest neighbours for
later use. In this paper, we take $N_{\rm HOP}=16$ as advocated by
\citet{EisensteinHut98}.
\item[(ii)] One locates the ``leaves'' of the tree, i.e. the most
elementary substructures, by associating groups of particles to local
SPH density maxima. This step is performed by a walk from particle to
particle, the next particle being the one with the maximum SPH density
among the particle itself and its $N_{\rm HOP}$ neighbours.
\item[(iii)] One can then establish the connectivity between these
``peak-patches'' by locating saddle points at the boundaries of the
above regions.
\item[(iv)] Finally, one builds the tree of structures and
substructures as a function of density threshold using the saddle
points to determine if two sub-structures are connected or not.
\end{itemize}

Note that we use a criterion relying on local Poisson noise in order
to assess a sub-structure's statistical significance: basically, a
structure of density $\rho$ and with $N$ particles must be at a
4-$\sigma$ level compared to local background, $\rho_{\rm b}$, to be
significant: $\rho > \rho_{\rm b} \times (1+4/\sqrt{N})$.

\subsection{Cross-match with FOF groups}

Our FOF haloes being identified with a linking-length parameter
$b=0.2$, we identify haloes with \ahop{} as connected regions of SPH
density larger than $\rho_{\rm th}=81$
\citep[e.g.][]{EisensteinHut98}. An additional fine-tuning of the
match between FOF and \ahop{} halo populations requires using $N_{\rm
SPH}=20$, in agreement with the minimum number of particles allowed in
FOF haloes. Still, the haloes produced by \ahop{} are unfortunately
slightly different from the FOF ones. We thus associate \ahop{}
sub-structures to FOF haloes according to one simple rule~: a
sub-structure is associated to the FOF halo which contains most of its
particles. This prescription has the advantage to avoid
mis-identifications, especially in rich environments, nearby massive
groups or clusters.

Also, because of mass resolution, nothing guarantees a priori that we
detect enough sub-structures to fit in all galaxies detectable by the
SDSS. We address this issue in Fig. \ref{fig:misidfracs} for galaxies
with $18<r<19$ (left-hand-side plot) and $19<r<20$ (right-hand-side
panel). The solid lines show the fraction of haloes containing more
detected galaxies than sub-structures in one of our mock
catalogues. Only haloes actually containing at least one galaxy were
considered for the normalisation of this fraction. The up-right
hatched region shows the red-shift distribution of haloes containing
at least one detected galaxy. Considering this region and the solid
curve, one sees that about $1\%$ of the haloes do not contain enough
sub-structures in both magnitude ranges. The dashed line shows the
fraction of galaxies which are not associated to sub-structures in the
same mock catalogue, that is, the fraction of galaxies that we have to
distribute on random halo particles. The red-shift distribution of
galaxies in the same catalogue is shown with an arbitrary
normalisation with the down-right hatches. Again, about $1\%$ of
detected galaxies only are not associated with sub-structures. These
two plots show that the level of contamination of the SUB scheme by
particles is at the $\sim$1\% level at most. In other words, the
clustering signal obtained with the SUB scheme does indeed come from
sub-structures.

\end{document}